\documentclass[aps,prb,twocolumn,superscriptaddress,showpacs]{revtex4-2}
\usepackage{graphicx} 
\usepackage{float}
\usepackage{geometry}
\usepackage{xcolor}
\usepackage{siunitx}

\usepackage{hyperref}
\hypersetup{
    colorlinks = true,
    citecolor = {blue},
    urlcolor = {blue}
}

\usepackage{color}
\definecolor{red}{rgb}{1,0,0}

\newcommand{\mathunit}[1]{~\mathrm{#1}}

\begin{document}

\title{Spin-lattice relaxation of NV centers in nanodiamonds adsorbed on conducting and non-conducting surfaces}

\author{Izidor Benedičič}
\affiliation{Jo\v{z}ef Stefan Institute, Jamova c. 39, 1000 Ljubljana, Slovenia}
\author{Yuri Tanuma}
\affiliation{Jo\v{z}ef Stefan Institute, Jamova c. 39, 1000 Ljubljana, Slovenia}
\author{Žiga Gosar}
\affiliation{Jo\v{z}ef Stefan Institute, Jamova c. 39, 1000 Ljubljana, Slovenia}
\affiliation{Faculty of Mathematics and Physics, University of Ljubljana, Jadranska c. 19, 1000 Ljubljana, Slovenia}
\author{Bastien An\'ezo}
\affiliation{Jo\v{z}ef Stefan Institute, Jamova c. 39, 1000 Ljubljana, Slovenia}
\affiliation{ Institut des Matériaux de Nantes Jean Rouxel (IMN), Nantes University, 2 Rue de la Houssinière, Nantes, France}
\author{Mariusz Mr\'ozek}
\affiliation{Faculty of Physics, Astronomy and Applied Computer Science, Jagiellonian University,  Prof. S. Łojasiewicza 11, 30-348 Krakow, Poland}
\author{Adam Wojciechowski}
\affiliation{Faculty of Physics, Astronomy and Applied Computer Science, Jagiellonian University,  Prof. S. Łojasiewicza 11, 30-348 Krakow, Poland}
\author{Denis Ar\v{c}on\thanks{Corresponding author.}}
\affiliation{Jo\v{z}ef Stefan Institute, Jamova c. 39, 1000 Ljubljana, Slovenia}
\affiliation{Faculty of Mathematics and Physics, University of Ljubljana, Jadranska c. 19, 1000 Ljubljana, Slovenia}
\email[Correspondence to:]{e-mail: denis.arcon@ijs.si}
\date{\today}

\begin{abstract}
   The nitrogen-vacancy color centers in nanodiamonds can be utilized as low-cost, highly versatile quantum sensors for studying surface properties in condensed matter physics through the application of relaxometry protocols. For such applications, a detailed knowledge of the intrinsic relaxation processes of NV centers in nanodiamonds is necessary. Here, we study the spin-lattice relaxation rates of NV ensembles in nanodiamonds with average diameters of 40~nm and 3~$\mu$m between room temperature and $\sim 6 \mathunit{K}$. 
   The NV relaxation curves fit to a stretched-exponential form with a stretching exponent $\alpha \approx 0.7$, implying the large distribution of relaxation times of individual centers within nanodiamonds.  
   We determine the Orbach-like scattering on phonons as the leading relaxation mechanism. 
   Finally, we discuss the viability of nanodiamonds as surface sensors when deposited on a metallic substrate and emphasize the need for well-controlled surface preparation techniques.
\end{abstract}

\maketitle

\section{Introduction}

\noindent
Spin-lattice relaxation rate $1/T_1$ is one of the most important parameters extensively measured in nuclear magnetic resonance experiments. $1/T_1$  contains the information on the spectrum of local magnetic field fluctuations and is thus sensitive to the lattice dynamics and the coupling to localized or itinerant spins \cite{abragamPrinciples1986}. In the former case, coupling to, e.g. lattice vibrations leads to Raman relaxation mechanisms, yielding characteristic power-law temperature dependencies of $1/T_1$. On the other hand, in metallic samples hyperfine coupling to itinerant charges gives rise to the so-called Korringa relaxation epitomized by $1/T_1T  = \mathrm {const}$, where $T_1$ is the spin-lattice relaxation rate and $T$ is the temperature \cite{korringaNuclear1950}.  However, nuclear magnetic resonance experiments require large quantities of the sample ($m \sim 100$~mg) due to the inherently low sensitivity of this experimental technique. Therefore, they cannot be applied in studies with small sample volumes, e.g., in studies of field fluctuations on the surface of materials \cite{walderOne2019}.   

A nitrogen-vacancy (NV) color center in a diamond is a point defect in the diamond lattice. It consists of a substitutional nitrogen atom (replacing a carbon atom) and a vacancy (missing carbon atom) in an adjacent lattice site. Recent advances in the detection and manipulation of NV centers have accelerated their use as quantum sensors in several fields, including magnetometry and relaxometry of solid-state systems \cite{ kumarRoom2024}, magnetic resonance detectors \cite{liuSurface2022} and temperature sensing \cite{wuHighsensitivity2022,dohertyTemperature2014}. As the excitation and detection of NV centers rely on optical methods, the NV relaxometry shows great improvement in sensitivity compared to nuclear magnetic resonance thereby opening the possibility of addressing local-field dynamics on surfaces. For example, measurements of $1/T_1$ of NV centers implanted at a depth of about $15\pm 10$~nm in a diamond substrate were employed to measure Johnson noise and ballistic transport in deposited silver films \cite{kolkowitzProbing2015}. Similarly, a single NV center in a scanning probe geometry has been employed for imaging of electrical conductivity at the nanoscale \cite{ariyaratneNanoscale2018}. 

Here we investigate the possibility of using NV centers in nanodiamonds (ND) to probe local field fluctuations on different surfaces. NDs are a system of particular interest in life sciences, where they can be used as non-bleaching, non-toxic sensors for monitoring ambient conditions or chemical reactions \cite{peronamartinezNanodiamond2020,petriniNanodiamond2022,neumannHighPrecision2013,kucskoNanometrescale2013}. 
On the other hand, the use of NV diamonds in nanoscale condensed matter studies remains largely unexplored.
This is due to the limited availability of commercially manufactured diamond probes for atomic force microscopy as well as the slow progress in the fabrication of diamond chips tailored to specific applications. Nevertheless, nanodiamonds offer numerous advantages. Due to their small size, implanted NV centers in nanodiamonds are always relatively close to the surface, a necessary condition for strong coupling to magnetic noise on the sample surface. Furthermore, one could in experiments distinguish between NV centers positioned on different magnetic domains, surface terminations, or nanofabricated structures. This provides spatial information about the local surface properties without the need for a scanning probe magnetometry setup. Combined with low costs and long relaxation times \cite{proothLong2023, marchLong2023}, nanodiamond-based relaxometry and magnetometry measurements could provide a versatile local alternative to bulk magnetometry and transport measurements.\par
For all the applications mentioned above, a good understanding of the intrinsic properties of nanodiamonds and their dependence on nanodiamond size is required. Despite this, the available data on the intrinsic spin-lattice relaxation in nanodiamonds is incomplete. The work focused on different sizes of nanodiamonds only looks at the relaxation at room temperature, while the studies of temperature-dependent properties mostly focus either on the low-temperature regime \cite{deguillebonTemperature2020} or the NV centers in bulk diamonds \cite{jarmolaTemperature2012, mrozekLongitudinal2015}. 
This work presents relaxometry measurements performed on NV center ensembles in nanodiamonds of various sizes deposited on conducting and insulating substrates. Specifically, we measure spin-lattice relaxation rates $1/T_1$  across a wide temperature range from room temperature to $6 \mathunit{K}$ and compare the relaxometry performance with relaxometry of single NV centers in diamond. 
We find that the sensitivity of the $T_1$ relaxometry for the local field fluctuations coming from the adsorbent   is limited by the significant distribution in the longitudinal relaxation rates and the tendency of smaller nanodiamonds to form agglomerates. Both factors  may limit the broader use of nanodiamond relaxometry in condensed matter physics problems.

\section{Experimental methods}

\begin{figure*}[htbp]
	\centering
	\includegraphics[width=\textwidth]{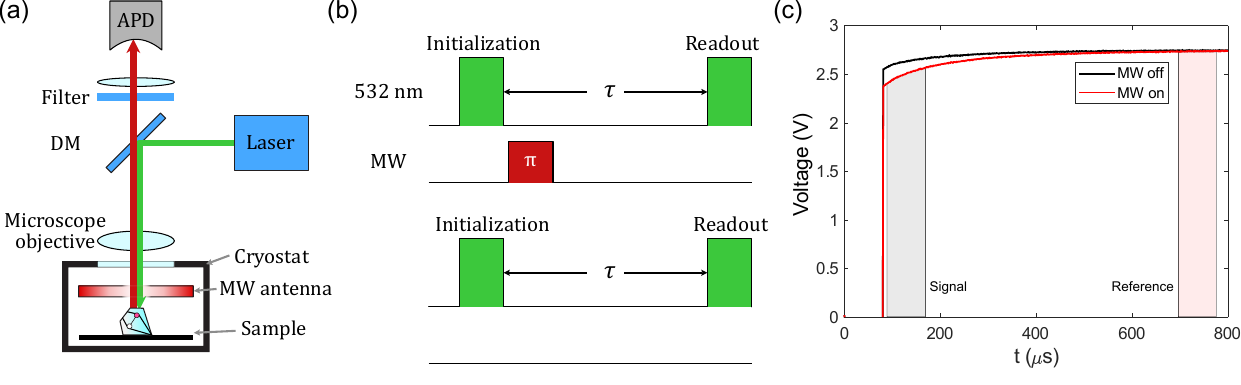}  
	\caption{(a) Schematic presentation of the experimental setup. The pump laser beam ($\lambda = 532 \mathunit{nm}$) is guided by a dichroic mirror (DM) and focused by a microscope lens on nanodiamonds, deposited on the Au or glass substrates. The focused laser light passes through an aperture in the microwave antenna, positioned just above the sample. The emitted light passes through a long-pass filter to eliminate the remaining laser green light and is collected on an avalanche photodiode (APD).
    (b) Pulse sequence for the $1/T_1$ relaxometry measurements. Subtracting the data without the microwave pulse from data gathered with the microwave pulse eliminates the signal contributions from light sources other than NV$^{-}$ centers.
    (c) Plot of a typical readout signal with (red curve) and without microwave (black curve) pulse. The two shaded areas mark the signal ($S$) and reference ($R$) regions of signal integration. In this experiment, the dark time is set to $\tau = 233$~$\mu$s.
    }
	\label{fig:setup}
\end{figure*}

For our experiments, we used fluorescent nanodiamonds purchased from Ad\'amas Nanotechnologies. The diamond particles of the target size are produced by milling the synthetic diamond manufactured by high-pressure high-temperature (HPHT) synthesis. They contain about 100 ppm of N substitutions and were irradiated with $2-3 \mathunit{MeV}$ electrons with fluence of $1\times10^{19} \mathunit{e/cm^2}$. The obtained samples were annealed at $850^{\circ} \mathrm{C}$ for 2 hour under vacuum \cite{adamasKnowledge,nunnBrilliant2019}. The diamond particles produced by this method  have a distribution of diameters \cite{adamasFunctionalized, reineckNot2019, ekimovSynthesis2023} and a significant anisotropy in the morphology, with shapes often resembling disks and rods instead of a sphere \cite{eldemrdashFluorescent2023}. The final concentration of NV centers was 1.5~ppm for nanodiamonds with an average diameter of 40~nm (on average around 14 NV centers per single nanodiamond) and 2.5~ppm for 3 $\mu$m diamonds (on average $8 \times 10^7$ NV centers per diamond). The nanodiamonds were dispersed in deionized water and deposited on the substrates using the drop-cast technique in a spin-coater rotating at 4000 RPM. High-purity thin films of gold on mica were purchased from Phasis S\`arl. For the glass substrate, we used microscope coverslips, cleaned in an ultrasonic bath with acetone and isopropanol. \par
The NV relaxometry measurements were performed with a home-built confocal microscopy setup, schematically presented in Fig. \ref{fig:setup} (a). All measurements were done in zero magnetic field, in a Lake Shore ST-500 continuous-flow cryostat cooled with liquid helium. The samples were mounted on a copper cold finger with a resistive heater, which enables measurements in a wide temperature range between room temperature and 5 K. A green laser beam with 532 nm wavelength (Coherent Verdi G5) was attenuated to 1.5 mW and focused on the sample using a microscope objective (Nikon CFI S Plan Fluor ELWD 20XC). From measurements of the collimated beam before the objective and the magnification, we estimate the diameter of the focused beam to be $d = 1 \mathunit{\mu m}$ in the beam waist. The light emanating from the sample was collected with the same objective lens and passed through a dichroic mirror (Thorlabs DMLP605) and a laser-line filter (Thorlabs NF533-17) to remove the residual laser green light. The red light was detected with a variable-gain avalanche photodiode (Thorlabs APD410A), and the photodiode signal was collected with an oscilloscope (Keysight DSOX1204G).
To deliver the microwave (MW) pulses to the sample, we used a home-built antenna, designed to allow for optical access while providing a homogeneous microwave magnetic field in the illuminated area. It consists of an omega-shaped copper stripline resonator on a printed circuit board (PCB), with the diameter of $d \sim 1 ~\mathrm{mm}$.
The microwave pulses were generated with a radiofrequency signal generator (Stanford Research Systems SG384). The laser and microwave pulses were programmed and synchronized using a SpinCore PulseBlasterESR-PRO.\par
Each measurement of $1/T_1$ consisted of the following steps. First, we measured the optically detected magnetic resonance (ODMR) spectrum in zero magnetic field. The signal was averaged from 64 to 1024 times, depending on the sample, to achieve a good signal-to-noise ratio. A Gaussian curve fitting was used to determine the average frequency of the resonance. In the next step, we measured Rabi oscillations and determined the optimal duration of the MW $\pi$ pulse.
Finally, we use the pulse sequence schematically shown in Fig. \ref{fig:setup} (b) to determine the $1/T_1$. The sequence begins with a $750 \mathunit{\mu s}$ long laser pulse, polarizing the NV centers to a $m_s = 0$ state. We then apply a MW $\pi$ pulse at the frequency $\nu_0$, corresponding to the average frequency of the resonance. A typical duration of the $\pi$ pulse is $\tau_{\pi} \sim 100 \mathunit{ns}$, as determined by the aforementioned measurement of Rabi oscillation. The $\pi$ pulse only excites the NV centers with the resonance frequency within the window of $1/2\tau_{\pi} = 5 \mathunit{MHz}$, much narrower than the full width of the spectrum, thus effectively selecting a sub-ensemble of NV centers. The system is then left to evolve for a variable time $\tau$, after which another laser pulse is applied to read the NV fluorescence and determine the residual spin polarization. The initial signal of the photodiode was integrated and normalized to the signal value after a long time, as illustrated in Fig. \ref{fig:setup} (c). The integration time intervals for signal ($S$) and reference ($R$) data were set to $80 \mathunit{\mu s}$. The normalized NV polarization signal is then calculated as $A = (R-S)/R$.
We then immediately apply a control sequence with the same parameters, except without the MW pulse. In the final stage the two results are subtracted to obtain a signal $A_\mathrm{CMR} = A_\mathrm{MW} - A_\mathrm{without~MW}$, which is proportional to the polarization of only those NV centers that were excited by the MW pulse \cite{jarmolaTemperature2012,mrozekLongitudinal2015}. This method thus eliminates the spurious signal from sources of photoluminescence other than the NV$^-$ centers excited by the MW pulse.


\section{Experimental results and discussion}

\begin{figure*}[htbp]
	\centering
	\includegraphics[width=\textwidth]{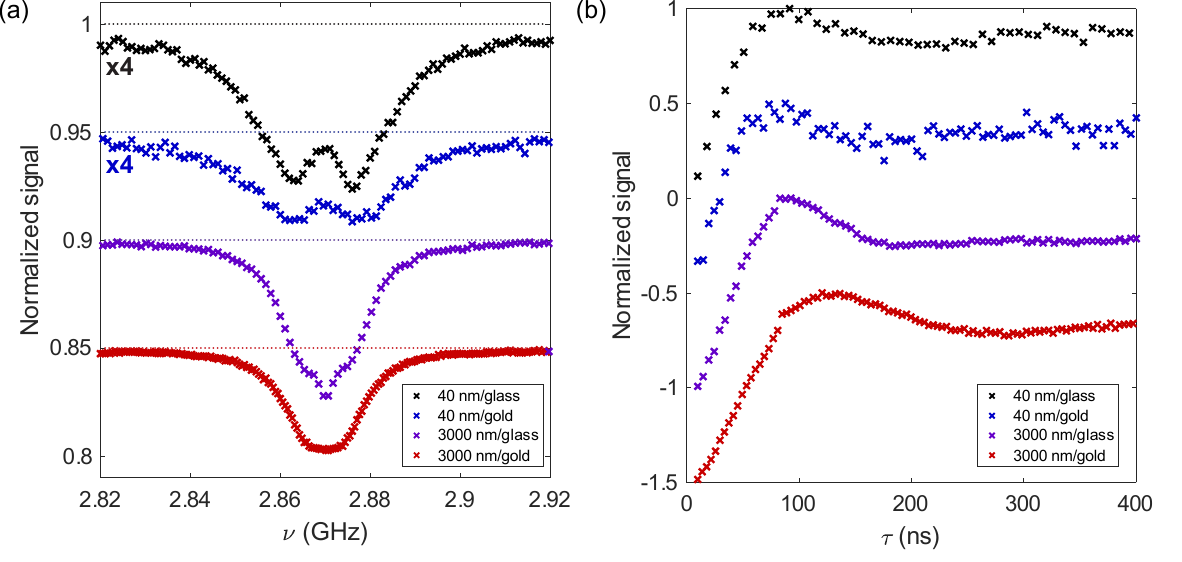}  
	\caption{(a) Typical ODMR signals for the four samples, taken at room temperature ($T = 295 \mathunit{K}$). Values are normalized to the baseline signal. Spectra taken on $40 \mathunit{nm}$ nanodiamonds are scaled by factor 4 for easy comparison. The curves are offset for clarity, dotted lines denote the baseline signal far from the resonance.
    (b) Measured Rabi oscillation curves for the spectra shown in panel (a). For clarity, curves are normalized to the maximum value of ODMR contrast. While the length of the optimal MW $\pi$ pulse varies slightly between different samples, most likely due to the variations in sample-MW antenna distances, we typically achieve a $\pi$ pulse length of $\tau_{\pi} = 100 \pm 20 ~\mathrm{ns}$.  
    In both panels, red, violet, blue and black crosses represent data collected on 3~$\mu$m NDs on Au, 3~$\mu$m NDs on glass, 40~nm NDs on Au and 40~nm NDs on glass substrates, respectively.
    }
	\label{fig:odmr}
\end{figure*}

The ODMR spectra of four different samples, 40~nm NDs deposited on Au and on glass substrates and 3~$\mu$m diamonds also deposited on Au and on glass, are compared in Fig. \ref{fig:odmr} (a). The samples with $3 \mathunit{\mu m}$ NDs exhibit a high contrast of $\sim 5 \%$, and the ODMR curve has a pronounced single central minimum at a MW frequency of $\nu_0= (2.870 \pm 0.005) \mathunit{GHz}$ at room temperature. The nanodiamonds with a smaller, $40 \mathunit{nm}$ diameter show a much weaker contrast of $\sim 1 \%$, and thereby worse signal-to-noise ratio (SNR). For these samples, we also notice a notable splitting of the ODMR minimum into two peaks separated by $\delta\nu = 13 \mathunit{MHz}$. The observed splitting does not originate from a magnetic field as the experiments were performed in a zero-field condition. Such splitting of the $m_S = \pm 1$ states can be attributed either to intrinsic strain \cite{alamDetermining2024} or a local electric field \cite{mittigaImaging2018}. In an ensemble of NV centers in several distinct nanodiamonds, each with slightly different strain levels, the strain distribution is expected to broaden the two-dip structure in a single wide minimum. The observed splitting is thus likely to originate from the local electric field, which leaves the two-dip structure intact in a much broader range of environmental conditions \cite{mittigaImaging2018}.

The Rabi oscillation measurements [Fig. \ref{fig:odmr} (b)] show in all cases a single strongly damped oscillation before settling the NV signal to a constant value. This indicates that in our measurements we do not observe a single NV center with a well-defined $m_S = 0 \rightarrow m_S = \pm 1$ transition, but rather an ensemble of centers with a distribution of resonant transitions. For short MW pulse durations, a broad maximum in Rabi oscillations is still visible, but the mixing of transitions for longer MW pulse durations makes the second maximum indistinguishable from the noise. The optimal pulse duration for an MW $\pi$ pulse is thus determined from the first Rabi oscillation as $\tau_\pi \approx 100 ~\mathrm{ns}$ for all four samples.

\begin{figure*}[htbp]
	\centering
	\includegraphics[width=\textwidth]{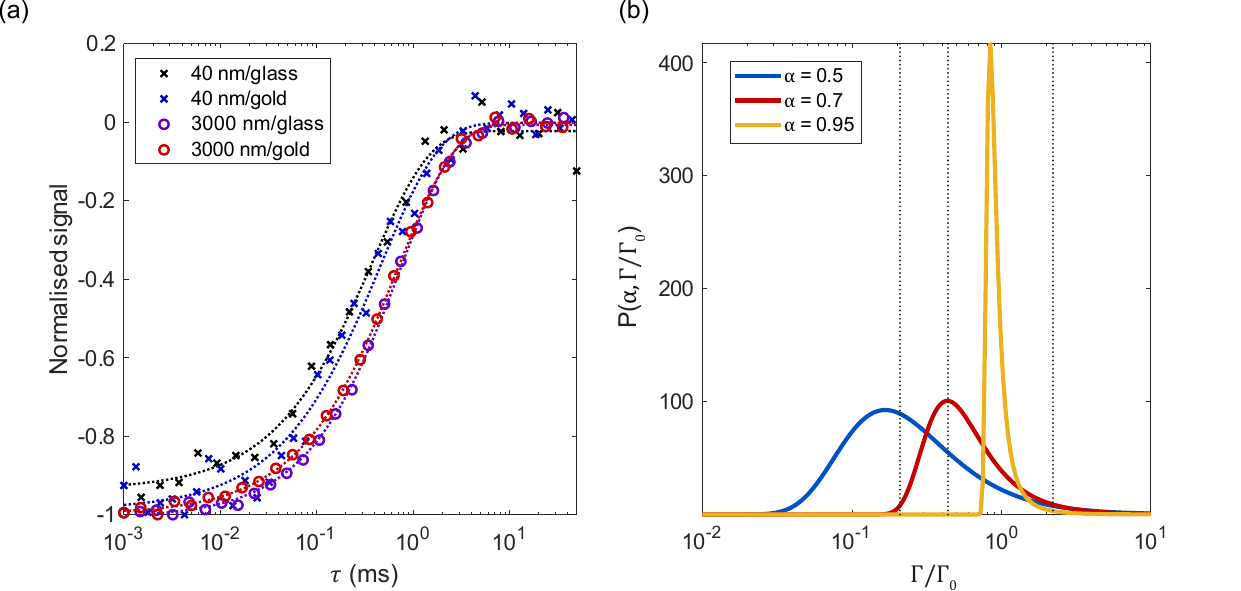}  
	\caption{(a) Characteristic room-temperature relaxation curves for four samples: NDs with 40~nm diameter on glass (black crosses), NDs with 40~nm diameter on Au (blue crosses), NDs with 3~$\mu$m  diameter on glass (red circles) and  NDs with 3~$\mu$m  diameter on Au (violet circles). Dotted lines are fits to Eq. \ref{eq:stretchedExp}. The characteristic relaxation times, $T_1$, for 3 $\mathrm{\mu m}$ diamonds are within the fitting uncertainty the same for both samples on glass and gold substrate ($740 \pm 20 ~\mathrm{\mu s}$ and $710 \pm 20 ~\mathrm{\mu s}$, respectively). With 40 nm diamonds, we see a reduction of $T_1$ by approximately $50 \%$, to $360 \pm 40 ~\mathrm{\mu s}$ and $460 \pm 40 ~\mathrm{\mu s}$ on glass and gold substrate, respectively. The stretching exponent $\alpha$ varies between $0.71 \pm 0.06$ and $0.76 \pm 0.02$.
    (b)
    Probability density distribution of relaxation rates $\Gamma$, normalized to the fitted characteristic relaxation time $\Gamma_0 = 1/T_1$. As $\alpha$ approaches unity, the distribution approaches the Dirac $\delta$ distribution (yellow line). For $\alpha < 1$, the distribution broadens and becomes asymmetric, thus $\Gamma_0$ cannot be interpreted as an average relaxation rate. For $\alpha = 0.7$ (red line), the vertical dotted lines denote the most likely relaxation rate (central line) and the values where the probability drops to 10 \% of the most likely value. 
    The distribution becomes even more spread out when $\alpha$ decreases, as demonstrated for the case of $\alpha= 0.5$ (blue line).
    }
	\label{fig:t1curve}
\end{figure*}

Next, we investigate the coupling of the NV centers in nanodiamonds to external magnetic noise using inversion-recovery relaxometry measurements schematically presented in Fig. \ref{fig:setup} (b). 
Examples of the NV magnetization relaxation curves taken at room temperature are shown in Fig. \ref{fig:t1curve} (a). The NV relaxation curves measured on $3 \mathunit{\mu m}$ diamonds are very similar for both substrates while relaxation curves show some dependence on the substrate for the smaller NDs with 40~nm diameter. 
The initial NV state, prepared with optical polarization and MW pulses, decays to a thermal mixture of $m_S = 0$ and $m_S = \pm 1$  states with a spin-lattice relaxation rate, $1/T_1$. 
We stress though that the signal cannot be described satisfactorily with a simple exponential function and instead has to be fitted with a stretched exponential function (also known as Kohlrausch function) of the form 
\begin{equation}\label{eq:stretchedExp}
    A = A_0 e^{ - \left( \tau/T_1 \right)^\alpha }.
\end{equation}
Here $A_0$ is a constant prefactor and $\tau$ is the dark time. The stretching parameter $\alpha$ empirically takes into account the distribution of $1/T_1$ values within the investigated spot. Therefore, the extracted relaxation rate parameter $1/T_1$ cannot be naively interpreted as an average relaxation rate,\cite{johnstonStretched2006} which will be later on important for further interpretation of the results. 

The fits to Eq. \ref{eq:stretchedExp} are plotted as dashed lines in Fig. \ref{fig:t1curve}. The extracted room-temperature relaxation times of NV centers in $3 \mathunit{\mu m}$ diamonds are $740 \pm 20 \mathunit{\mu s}$ and $710 \pm 20 ~\mathrm{\mu s}$ for the glass and gold substrates, respectively.
By comparison, the relaxation of NV centers in $40 \mathunit{nm}$ NDs deposited on the same two substrates is noticeably faster [black and blue curves in Fig. \ref{fig:t1curve} (a)], with $T_1$ reduced down to  $360 \pm 40 ~\mathrm{\mu s}$ and $460 \pm 40 ~\mathrm{\mu s}$, respectively. 
The stretching exponent $\alpha$ is between $0.71 \pm 0.06$ and $0.76 \pm 0.02$ for all four samples.
The shortening of relaxation time is consistent with on average shorter distance between the NV centers and the surface paramagnetic defects on individual nanodiamonds \cite{tetienneSpin2013}.
However, the two values for NDs on different substrates are still rather close and the extracted $T_1$ of nanodiamonds on glass is even shorter than that for the nanodiamonds deposited on gold.\par

The stretched exponential relaxation behavior is ubiquitous yet rarely discussed in NV relaxometry. While single NV emitters in nanodiamonds are known to relax with a simple exponential curve \cite{deguillebonTemperature2020}, deviations from exponential relaxation were observed both in diamond nanopillars, where the stretching exponent $\alpha$ was temperature independent and attributed to the distribution of intrinsic relaxation times \cite{kumarRoom2024}, and in biological systems \cite{peronamartinezNanodiamond2020}. In some instances, it is possible to approximate the NV magnetization relaxation curve with a two-exponential model \cite{kumarExcitation2016, peronamartinezNanodiamond2020}, where a slowly relaxing component corresponds to NV centers far away from the nanodiamond surface and a fast relaxing component is attributed to NV centers closer to the surface where the relaxation effects of paramagnetic impurities are stronger. However, this approach is hard to implement in our case, where the NV relaxation curves do not clearly show two distinct time scales, and attempts to fit with a two-exponential curve result in a very long slow component even on the smallest nanodiamonds, inconsistent with the known properties of nanodiamonds \cite{tetienneSpin2013} (see Fig. S1 in Supplementary material for comparison of different fit functions \cite{supplementaryMaterial}).
Thus, we use the stretched exponential curve to fit the data instead. 

As stated above, the stretched exponential relaxation curve can be interpreted as emerging from a distribution of relaxation rates across various timescales \cite{kohlrauschTheorie1854}. Mathematically, this can  be expressed as
\begin{equation}
    e^{ - \left( \tau/T_1 \right)^\alpha } = \int_0^\infty P(s,\alpha) e^{-s \tau/T_1} ds,
\end{equation}
where $T_1$ is the characteristic spin-lattice relaxation time, and $s = T_1/T_1^\mathrm{local} =  \Gamma/\Gamma_0$ is the dimensionless relaxation rate parameter. Here, $\Gamma_0 = 1/T_1$ is the characteristic parameter obtained from fits to Eq. \ref{eq:stretchedExp}, and the local relaxation rate $\Gamma = 1/T_1^\mathrm{local}$ is a property of a particular NV center that contributes to the total signal. The distribution of local rates $\Gamma$ is expressed with the stretching exponent  $\alpha$ as 
\begin{equation}
    P(\alpha, s) = \int_{-\infty}^\infty e^{(-i u)^\alpha}e^{isu}du,
\end{equation}
where $i$ is the imaginary unit. The calculated distribution curves for some characteristic values of $\alpha$ are compared in Fig. \ref{fig:t1curve} (b): for $\alpha \rightarrow 1$, the distribution approaches a Dirac $\delta$ distribution, corresponding to a single well-defined value of $1/T_1$. For $\alpha<1$, the distribution becomes broadened and skewed, with the relaxation rate $\Gamma_{\rm max} < \Gamma_0$ at which $P(\alpha, s)$ has a maximum and a long tail towards $\Gamma > \Gamma_0$. The fitting parameter $1/T_1$ is therefore not an average but a median value of the distribution \cite{johnstonStretched2006}.

To obtain a quantitative estimation of the spread of relaxation rates for our system, we have numerically calculated the distribution $P(\alpha,s)$ for $\alpha = 0.7$, a value close to the extracted parameters from the fits to all four curves in Fig. \ref{fig:t1curve}. The distribution in $1/T_1$ becomes in this case already very broad, extending over more than a decade in the $\Gamma /\Gamma_0$ ratio. The maximum in $P(\alpha,s)$ is found at  $1/T_1^{max} = 0.44/T_1$  [central vertical line in Fig. \ref{fig:t1curve} (b)]. The interval within which the probability density drops to 10 \% of that at $T_1^{max}$ (dotted vertical lines in Fig. \ref{fig:t1curve} (b)) spreads between the lower and upper boundaries of $0.2/T_1$ and $2.2/T_1$, respectively. The distribution in relaxation rates of NV centers in NDs thus spans over more than an order of magnitude, and the extracted relaxation parameter $1/T_1$ overestimates the most likely single relaxation rate by approximately a factor of 2.
While the differences in the NV relaxation curves between different samples are significant, the values of $\alpha$ remain similar, and thus the single relaxation parameter $1/T_1$ extracted from the NV relaxation curve fits to Eq. \ref{eq:stretchedExp} can still be a useful figure of merit as long as $\alpha$ remains close to unity. 

\begin{figure*}[ht]
	\centering
	\includegraphics[width=\textwidth]{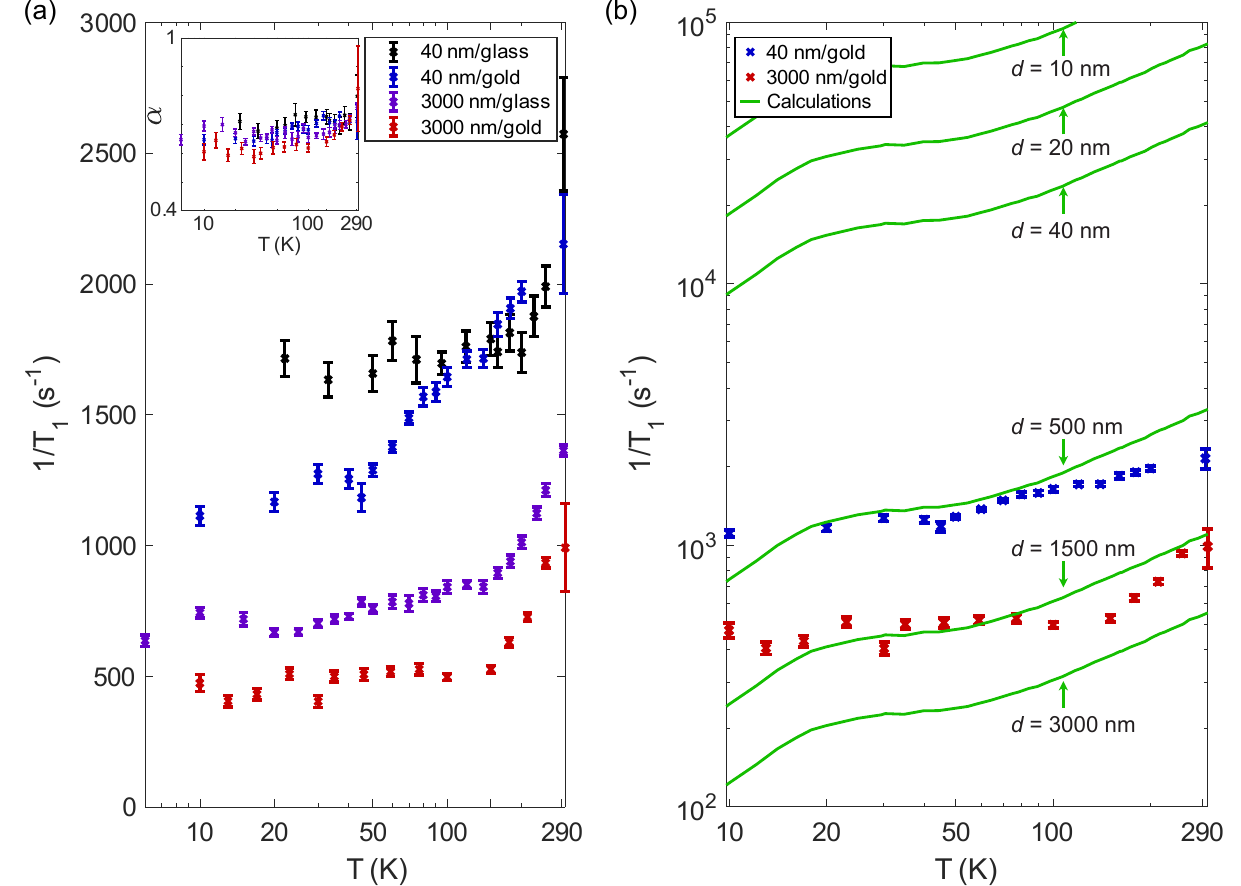}  
	\caption{ (a) Temperature dependence of $1/T_1$ for the four measured samples: NDs with $40 \mathunit{nm}$ diameter on glass (black), NDs with $40 \mathunit{nm}$ diameter on Au (blue), diamonds with $3 \mathunit{\mu m}$ diameter deposited on glass (violet) and on Au (red). In three of the samples, the relaxation rate initially decreases upon cooling and saturates below $T \sim 150 ~\mathrm{K}$. The exception is the sample with $40 \mathunit{nm}$ nanodiamonds deposited on Au, where the rate keeps decreasing until $T \sim 50 ~\mathrm{K}$.
    Insert: stretching exponent, $\alpha$, as a function of temperature for all four samples.
    (b) Comparison of experimental relaxation rate for $40 \mathunit{nm}$ NDs (blue) and $3 \mathunit{\mu m}$ diamonds (red) on a gold substrate with calculated relaxation contributions from Johnson noise (green lines). The green curves were calculated for several fixed distances between the NV centers and the substrate (see labels next to the lines).
    }
	\label{fig:tdep}
\end{figure*}

Next, we discuss the temperature-dependent NV relaxometry results. The temperature dependencies of parameters $1/T_1$ and $\alpha$ for all four samples are summarized in Fig. \ref{fig:tdep} (a). In all cases, $1/T_1$ shows a qualitatively similar temperature dependence: the relaxation rate is the largest at high temperatures and rapidly decreases upon cooling. The $1/T_1$ values at low temperatures approach a constant value that also depends on the particular spot of measurement. The temperature below which $1/T_1$ becomes approximately temperature-independent varies between approximately $50 \mathunit{K}$ and $200 \mathunit{K}$ and depends on the sample but to a certain degree also on the investigated spot. 

The temperature dependence of the NV relaxation rate measured in this work is to be compared with the data for NV centers in nanodiamonds published in Ref. \cite{deguillebonTemperature2020}. There, the measurements suggested an almost temperature-independent relaxation rate even at high temperatures. The reported relaxation rates, $1/T_1 > 3 \cdot 10^3 \mathunit{s^{-1}}$ are significantly higher than in the present measurements, by about one order of magnitude at the highest temperatures. However, we find the ratio between the relaxation rates at low and high temperatures, $(1/T^\mathrm{10 \mathunit{K}}_1)/(1/T^\mathrm{RT}_1)$, is very close for all samples, with the average ratio $(1/T^\mathrm{10 \mathunit{K}}_1)/(1/T^\mathrm{RT}_1) = 0.5 \pm 0.1$ matching the previously reported value of $0.41 \pm 0.1$ within the uncertainty interval \cite{deguillebonTemperature2020}.
The data of Ref. \cite{deguillebonTemperature2020} also showed a downturn at very low temperatures ($T < 20 \mathunit{K}$), which was suggested to be a signature of a thermally activated relaxation process. We do not find such a downturn in our data, where the lowest measured temperature is $10 \mathunit{K}$. 


In general, the NV relaxation rate, $1/T_1$, has two independent contributions
\begin{equation}\label{eq:relaxRateGeneral}
    \frac{1}{T_1} = \left( \frac{1}{T_1} \right)_{\rm int} + \left( \frac{1}{T_1} \right)_{\rm sub}\quad .
\end{equation}
The first term, $\left(1/T_1 \right)_{\rm int}$, is governed by the intrinsic relaxation mechanisms of the nanodiamonds, while the second term, $\left(1/T_1 \right)_{\rm sub}$, is due to the fluctuations of local field originating from the glass/Au substrate. The latter is expected to be negligible for the glass substrate. In the following paragraphs, we first discuss the measurements performed on glass substrates before attempting to extract $\left(1/T_1 \right)_{\rm sub}$ for the metallic Au substrates.

To determine $\left(1/T_1 \right)_{\rm int}$, we first note that the observed dependence of $1/T_1$ is similar to the NV relaxation in bulk diamonds, where the phononic relaxation mechanisms are suppressed at low temperatures \cite{mrozekLongitudinal2015}. While some studies include the two-phonon Raman term proportional to $T^5$ at high temperatures, we find this term negligible in the temperature range of present measurements. This is consistent with recent calculations suggesting the two-phonon Raman relaxation is negligible even at temperatures above $1000 \mathunit{K}$ \cite{cambriaTemperatureDependent2023}. 
Therefore, we fit the $1/T_1$ dependencies with an expression commonly employed for NV centers in bulk diamonds
\begin{equation}\label{eq:t1relaxation}
    1/T_1 = a_1 + \frac{a_2}{e^{\Delta/k_{\rm B}T} - 1}.
\end{equation}
Here the first, temperature-independent term $a_1$ originates in the relaxation due to dipolar interactions with the diluted moments in the NDs. The second, Orbach-like phonon relaxation term with a characteristic phonon energy $\Delta$ is due to the spin-lattice relaxation, where the Raman scattering is driven by the second-order interactions \cite{cambriaTemperatureDependent2023}.

Fitting of the temperature dependence of $1/T_1$ [Fig. \ref{fig:tdep} (a)] to Eq. \ref{eq:t1relaxation}  yields parameters summarized in Table \ref{tab:fitparams}. 
\begin{table}
    \centering
    \begin{tabular}{|c || c c c|} 
        \hline
        & $a_1$ (s$^{-1}$) & $a_2$ (s$^{-1}$) & $\Delta$ (meV)  \\
        \hline\hline
        ND40/glass & 1730 & $3 \times 10^{5}$ & 150  \\ 
        \hline
        ND40/gold & 1150 & 240 & $<$ 5  \\
        \hline
        ND3000/glass & 720 & 1400 & 30  \\
        \hline
        ND3000/gold & 480 & 1150 & 30 \\
        \hline
    \end{tabular}
    \caption{Fitting parameters of the temperature dependencies of $1/T_1$ for the model described with Eq. \ref{eq:t1relaxation}.}
    \label{tab:fitparams}
\end{table}
\noindent
The parameter $a_1$, corresponding to the low-temperature saturation value, is enhanced by more than one order of magnitude compared to the values in the range between $0.07 \mathunit{s^{-1}}$ and $21 \mathunit{s^{-1}}$ reported for the bulk diamonds \cite{jarmolaTemperature2012,mrozekLongitudinal2015}. In contrast to the low-temperature relaxation of NV centers in bulk diamonds, where $a_1$ is interpreted as cross-relaxation between different NV centers, the value of $a_1$ in nanodiamonds is governed by a dipolar coupling to paramagnetic defects on the surface. This is further corroborated by the marked difference between the $40 \mathunit{nm}$ and $3 \mathunit{\mu m}$ diameter diamonds: nanodiamonds show higher relaxation rates, consistent with a larger surface-to-volume ratio of smaller particles.\par

The second term in Eq. \ref{eq:t1relaxation} has two free parameters, the constant $a_2$ and the characteristic phonon energy $\Delta$, determining the temperature at which the relaxation via phononic processes becomes important. 
{\em Ab initio} calculations show that phonon modes that change the positions of the carbon dangling bonds display two peaks in the spectrum of local field fluctuations at around $65 \mathunit{meV}$ and $155 \mathunit{meV}$, which sets the scale for the phonon energy  $\Delta$ \cite{cambriaTemperatureDependent2023}.
As is evident from Fig. \ref{fig:tdep} and the fitting parameters (Table \ref{tab:fitparams}), the extracted values of $\Delta$ are close to these calculated phonon energies, except for the notable case of 40~nm NDs deposited on Au and which will be discussed later. The high-temperature relaxation mechanism probed in our experiments is thus most likely Raman scattering driven by the second-order interactions introduced in Ref. \cite{cambriaTemperatureDependent2023}.

We can summarize at this point that the temperature dependence of intrinsic NV relaxation rates in diamonds with micrometer and nanometer sizes down to $d = 40 \mathunit{nm}$ is qualitatively very similar to that for NV centers in the bulk diamond \cite{jarmolaTemperature2012,rosskopfInvestigation2014}, but with a substantially enhanced temperature-independent dipolar relaxation contribution marking the significant contributions from the paramagnetic impurities at the ND surface. While we do not find any evidence of a low-temperature thermally activated relaxation process similar to that in Ref. \cite{deguillebonTemperature2020}, we cannot exclude the possibility that the activation energy in our system is so low that its contribution is already saturated at $T = 10 \mathunit{K}$.

Turning next to the temperature dependence of the stretching exponent $\alpha$, we find it is almost independent of temperature and does not significantly change with diamond particle size or substrate type [insert to Fig. \ref{fig:tdep} (a)].
The very weak temperature dependence of $\alpha$ appears correlated with the trend observed in the relaxation rate $1/T_1$: when the relaxation rate decreases with decreasing temperature, the value of $\alpha$ also decreases. However, the changes are small, ranging from $\alpha \approx 0.75$ at room temperature to $\alpha \approx 0.65$ at lowest temperatures. The tendency of $\alpha$ to approach 1 at higher temperatures may be due to the more efficient spin diffusion processes at elevated temperatures and thus the more efficient establishment of common spin temperature within a single ND.  \par
Although it is commonly observed \cite{kumarRoom2024, vedelaarOptimizing2023, reineckNot2019}, the origin of the $1/T_1$ distribution in NV relaxometry is still inadequately understood. Several aspects need to be considered.
One possibility is that the inhomogeneity of magnetic fluctuations across the illuminated substrate surface results in variations in $1/T_1$ in different nanodiamonds. If several nanodiamonds are illuminated at once, this would result in the observed stretched exponential distribution. 
This, however, is unlikely because the stretching exponent is almost the same for samples with different sizes of nanodiamonds. While the laser beam illuminates several $40 \mathunit{nm}$ diamonds simultaneously, the focused beam diameter is comparable to the size of a single $3 \mathunit{\mu m}$ diamond, making the illumination of several microdiamonds improbable.
Therefore, we conclude that the value of $\alpha$ results from the intrinsic random distribution of the NV centers within the nanodiamonds. As reported in Ref. \cite{tetienneSpin2013}, the precise position of the NV center within a nanodiamond strongly influences the dipolar relaxation to the paramagnetic defects on the diamond surface. In particular, the difference between the lower and the upper bound is as much as two orders of magnitude (from $10^{3} \mathunit{s^{-1}}$ to $10^5 \mathunit{s^{-1}}$ according to the assumptions of the mentioned work) for diamonds larger than $\sim 30 ~\mathrm{nm}$ in diameter.
In our experiment, the nanodiamonds host several NV centers per particle, on average 14 NV centers per single $40 \mathunit{nm}$ nanodiamond. Due to the aggregation of nanodiamonds, it is possible to illuminate several nanodiamonds at once, further increasing the number of excited NV centers. Additionally, not all nanodiamonds can be assumed to be equally sized and strained due to randomness in the manufacturing process \cite{finasHow2024}. All these contributions would lead to a stretched exponential relaxation curve, where the distribution of $1/T_1$ originates from the distribution of $1/T_1$ rates of the individual NV centers within a particle. This also explains why in nanodiamonds with a single NV center a single-exponential behavior has been observed \cite{deguillebonTemperature2020}. \par

Finally, we discuss the contribution of the substrate to the total relaxation rate of deposited NDs.
The spin-lattice relaxation rate $1/T_1$ of $40 \mathunit{nm}$ NDs deposited on Au substrate [blue curve in Fig. \ref{fig:tdep} (a)] with a fit to Eq. \ref{eq:t1relaxation} yielding anomalously small $\Delta < 5 \mathunit{meV}$ and $a_2$ levels off only at a relatively low temperature of $T \approx 50 \mathunit{K}$. This could indicate significant coupling of NV centers in the $40 \mathunit{nm}$ nanodiamonds to the magnetic field fluctuations of the conductive surface. \par
In general, the relaxation rate of an NV center due to the fluctuations of magnetic fields originating from the substrate can be written as
\begin{equation}\label{eq:relaxRateGeneral}
    \left( \frac{1}{T_1} \right)_{\rm sub} = \int \gamma^2 \langle B^2 \rangle S (\omega, T) F(\omega ) d\omega,
\end{equation}
where $\gamma$ is the gyromagnetic ratio of the NV center, and $\sqrt{\langle B^2 \rangle }$ effective magnetic field at the position of the NV center. $S(\omega)$ is the spectral density of magnetic noise, and $F(\omega)$ is the filter function, which depends on the utilized pulse sequence. For an inversion-recovery $1/T_1$ measurement, the filter can be approximated as a narrow window centered at the ODMR frequency $\omega \approx 2\pi \cdot 2.87 \mathunit{GHz}$ at room temperature \cite{schafer-nolteTracking2014}. In the presence of an external noise source, such as the Johnson-Nyquist noise in a good conductor, the relaxation rate $1/T_1$ is increased proportionally to the conductivity of the sample \cite{kolkowitzProbing2015,ariyaratneNanoscale2018}. Specifically, the relaxation rate  from the Johnson noise of the conducting substrate is of the form
\begin{equation}
    \left( \frac{1}{T_1} \right)_{\rm sub} \propto T \sigma(T),
\end{equation}
where $T$ is the sample temperature and $\sigma(T)$ the temperature-dependent electrical conductivity of the sample. Following Kolkowitz \textit{et al.} \cite{kolkowitzProbing2015}, we calculate the expected temperature dependence of $(1/T_1)_{\rm sub}$ for the Au substrate for several distances $d$ between the NV center and substrate, ranging from $d = 10 \mathunit{nm}$ to $d = 3000 \mathunit{nm}$. We plot the calculated curves in Fig. \ref{fig:tdep} (b) together with our experimental data. 

Assuming the average distance between the NV center and the substrate is half of the ND diameter, the calculated relaxation rate is more than an order of magnitude larger than what we observe in the experiment taken on NDs with $40 \mathunit{nm}$ diameter [blue points in Fig. \ref{fig:t1curve}(b)]. Even accounting for the broad distribution of $1/T_1$ values of individual NV centers, the theoretical calculations suggest a much stronger relaxation rate than observed.
Only for NV centers at a significantly larger distance of $d > 500 \mathunit{nm}$ from the conductive substrate, the $(1/T_1)_\mathrm{sub}$ contribution would become comparable in magnitude to the observed total $1/T_1$ values.
This observation may suggest the aggregation of $40 \mathunit{nm}$ nanodiamonds as a cause for the lack of significant coupling to current noise from the conductive substrate. Namely, in the experiment the spots with larger ND agglomerates yield stronger signals, thereby dominating the observed response.
Furthermore, in an ensemble with several NDs, the centers with the strongest coupling to the environment will have the shortest $T_1$, on the order of $\sim 100 ~\mu\mathrm{s}$. The signal originating from these centers will quickly vanish and the relaxation curve will be dominated by the more distant NVs with longer $T_1$.
The same effect could also be responsible for the apparently faster relaxation of NDs on glass substrate compared to those deposited on gold. The $T_1$ of NV centers strongly coupled to the substrate could become immeasurably short, leaving the NDs further away from the substrate dominating the signal. However, since we do not observe any significant change in $\alpha$ between the substrates, this mechanism cannot be unambiguously confirmed. \par

For larger diamonds with $3 \mathunit{\mu m}$ diameter [red points in Fig. \ref{fig:t1curve}(b)], the calculated contribution of $(1/T_1)_{\rm sub}$ is approximately the same as the measured $1/T_1$. However, since the relaxation rates measured on gold are very close to the ones measured on glass, and because the $1/T_1$ values of the individual NVs can vary by one order of magnitude (indicated by the value of stretching exponent $\alpha \approx 0.7$), one cannot claim the $3 \mathunit{\mu m}$ ND show any significant coupling to the magnetic noise of the substrate.

The deposition of diamonds with larger diameters is straightforward, as samples prepared with a simple drop-cast technique show good surface coverage of well-separated single diamond particles (Fig. S2 in Supplementary material \cite{supplementaryMaterial}). However, our measurements indicate that diamonds so large are not sensitive enough to be used for detecting small variations in surface properties. Smaller, $40 \mathunit{nm}$ NDs would be a better candidate, but are more prone to forming aggregates, which then dominate the signal.\par
\begin{figure}[ht]
	\centering
	\includegraphics[width=0.5\textwidth]{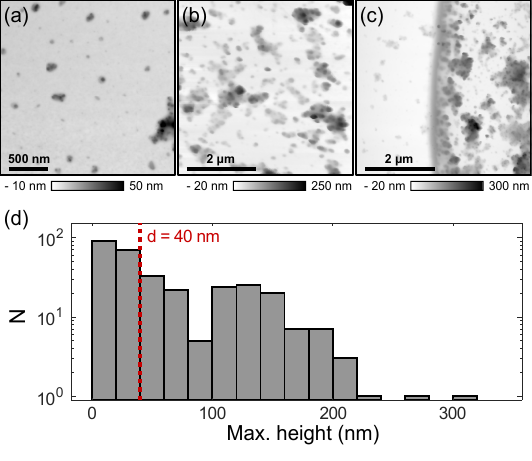}  
	\caption{ (a)-(c) Typical AFM images of the sample of diamond particles with a nominal size of 40 nm, deposited on glass. The images were acquired on different parts of the same sample.
    (d) Histogram of maximal cluster heights obtained from the images in panels (a)-(c). While the most common heights are lower or comparable to the nominal diamond particle diameter $h \approx d$, marked with red dashed line, larger clusters can reach heights of several hundreds nm.
    }
	\label{fig:clusters}
\end{figure}

To quantify the aggregation of the nanodiamonds used in our experiment, the surface topography of one of the samples with $40 \mathunit{nm}$ diamonds deposited on glass substrate is next investigated with atomic force microscopy (AFM).
Some typical AFM images are shown in Fig. \ref{fig:clusters} (a)-(c), where one can see areas with individual nanodiamonds and small clusters (Fig. \ref{fig:clusters} (a)) as well as areas where significant ND aggregation took place during the deposition (Fig. \ref{fig:clusters} (b), (c)). The typical distance between the aggregates is smaller than the estimated laser beam diameter of $d_{\rm laser} \approx 1 \mathunit{\mu m}$.
The analysis of cluster heights, Fig. \ref{fig:clusters} (d), shows that most observed diamond particles are smaller than the nominal size, $h < d = 40 \mathunit{nm}$. This reflects the previously reported observations that milled nanodiamonds often form anisotropic, disk-shaped particles \cite{eldemrdashFluorescent2023}.
While clusters with large maximum heights are infrequent, approximately $3\%$ of the clusters are higher than $h = 200 \mathunit{nm}$, and the NV centers in such clusters would be only very weakly coupled to the magnetic fluctuations originating in the substrate. Combined with small inter-cluster spacing, these anomalous aggregates are likely to dominate the detected NV signal at a given spot, resulting in an apparent insensitivity to the choice of the substrate. \par
The aggregation of nanodiamonds dispersed in a liquid medium is commonly observed, especially in studies of detonation nanodiamonds, where diameters of the observed diamond clusters can be as much as an order of magnitude larger than the diameters of individual nanodiamonds \cite{kuschnerusFabrication2023,katsievFresh2021,bolshedvorskiiSingle2017}. While the aggregation of HPHT-grown nanodiamonds is less dramatic \cite{zeleneevStudying2020}, it is nevertheless always present. In addition, the commercially purchased nanodiamonds are not of uniform size but follow a relatively broad size distribution, peaked at the nominal diameter of $d = 40 \mathunit{nm}$. Looking at the available DLS data \cite{adamasFunctionalized} reveals a notable fraction of nanoparticles with diameters $d > 100 \mathunit{nm}$. A combined effect of particle aggregation and the presence of larger diamond particles could result in an increased average distance between the NV centers and the substrate. However, the shape asymmetry of the particles \cite{eldemrdashFluorescent2023} further complicates this picture. The widespread use of nanodiamonds as surface sensors thus critically depends on careful sample preparation. One possible option would be to use a deposition technique that does not include the deposition of a dispersion medium, which facilitates the diffusion and aggregation of the NDs.

\section{Conclusions}

The temperature dependence of spin-lattice relaxation rates of NV centers in nanodiamonds shows a rapid decrease upon cooling from room temperature, consistent with the phononic relaxation mechanisms observed with NV centers in bulk diamonds. While expected in larger, 3 $\mu$m diamonds, this behavior persists even in 40 nm nanodiamonds. Previous studies have focused either on low-temperature \cite{deguillebonTemperature2020} or high-temperature \cite{cardosobarbosaTemperature2024, liuCoherent2019} properties of NV centers in nanodiamonds, while the intermediate temperature regime remained poorly explored yet it is of great interest for the solid-state applications.
All investigated samples show a stretched exponential relaxation behavior, indicating a significant distribution of local relaxation rates of NV centers within nanodiamonds. The extracted stretching-exponent parameters $\alpha \approx 0.7$ are almost temperature-independent with a slight increase at higher temperatures.
The relaxation-rate parameter $1/T_1$ obtained from the fits of the relaxation curves must therefore be interpreted as a median relaxation rate of the NV centers within an ensemble. 

Upon cooling from room temperature, the values of $1/T_1$ initially decrease and then saturate below $\sim 150 ~\mathrm{K}$, depending on the sample. This indicates that the high-temperature relaxation in nanodiamonds is dominated by the Orbach-like phonon process, as is the case for NVs in bulk diamonds. The low-temperature saturation value is significantly enhanced compared to NVs in bulk diamonds due to enhanced coupling to the surface paramagnetic defects.
We do not find convincing signatures for the coupling of NV centers to the Johnson noise of the metallic substrates, neither in the magnitude nor in the temperature dependence of the relaxation rate $1/T_1$. Comparing the measured data with the calculated contribution of the Johnson noise to the spin-lattice relaxation rate, we find the sensitivity of $40 \mathunit{nm}$ diamonds is likely limited by the nanoparticle aggregation, while the intrinsic spread of $1/T_1$ values within the ensemble is limiting the use of larger, $3 \mathunit{\mu m}$ diamonds.

\begin{acknowledgments}
The authors thank dr. Miha Škarabot for acquiring the AFM images.
This research was funded in part by the National Science Centre, Poland grant 2020/39/I/ST3/02322 and grant 2023/05/Y/ST3/00135 within the QuantERA II Programme that has received funding from the European Union’s Horizon 2020 research and innovation program under Grant Agreement No 101017733.
The research was also supported by the Slovenian Research Agency under project grant N1-0220 and the research program P1-0125.
Y. T. acknowledges financial support by the Japanese "JSPS Overseas Research Fellowships" scheme.
For the purpose of Open Access, the author has applied a CC-BY public copyright license to any Author Accepted Manuscript (AAM) version arising from this submission. \\
The research data is available at \cite{benedičič_gosar_mrózek_wojciechowski_tanuma_arčon_anézo_2025}.
\end{acknowledgments}

\bibliography{nanodiamondsPaper}

\end{document}


\title{Supplementary information for: Spin-lattice relaxation of NV centers in nanodiamonds adsorbed on conducting and non-conducting surfaces}

\author{Izidor Benedičič}
\affiliation{Jo\v{z}ef Stefan Institute, Jamova c. 39, 1000 Ljubljana, Slovenia}
\author{Yuri Tanuma}
\affiliation{Jo\v{z}ef Stefan Institute, Jamova c. 39, 1000 Ljubljana, Slovenia}
\author{Žiga Gosar}
\affiliation{Jo\v{z}ef Stefan Institute, Jamova c. 39, 1000 Ljubljana, Slovenia}
\affiliation{Faculty of Mathematics and Physics, University of Ljubljana, Jadranska c. 19, 1000 Ljubljana, Slovenia}
\author{Bastien An\'ezo}
\affiliation{Jo\v{z}ef Stefan Institute, Jamova c. 39, 1000 Ljubljana, Slovenia}
\affiliation{ Institut des Matériaux de Nantes Jean Rouxel (IMN), Nantes University, 2 Rue de la Houssinière, Nantes, France}
\author{Mariusz Mr\'ozek}
\affiliation{Faculty of Physics, Astronomy and Applied Computer Science, Jagiellonian University,  Prof. S.Łojasiewicza 11, 30-348 Krakow, Poland}
\author{Adam Wojciechowski}
\affiliation{Faculty of Physics, Astronomy and Applied Computer Science, Jagiellonian University,  Łojasiewicza 11, 30-348 Krakow, Poland}
\author{Denis Ar\v{c}on\thanks{Corresponding author.}}
\affiliation{Jo\v{z}ef Stefan Institute, Jamova c. 39, 1000 Ljubljana, Slovenia}
\affiliation{Faculty of Mathematics and Physics, University of Ljubljana, Jadranska c. 19, 1000 Ljubljana, Slovenia}
\email[Correspondence to:]{e-mail: denis.arcon@ijs.si}
\date{\today}

\maketitle

\section{Optical images of samples}

\begin{figure*}[htbp]
	\centering
	\includegraphics[width=0.8\textwidth]{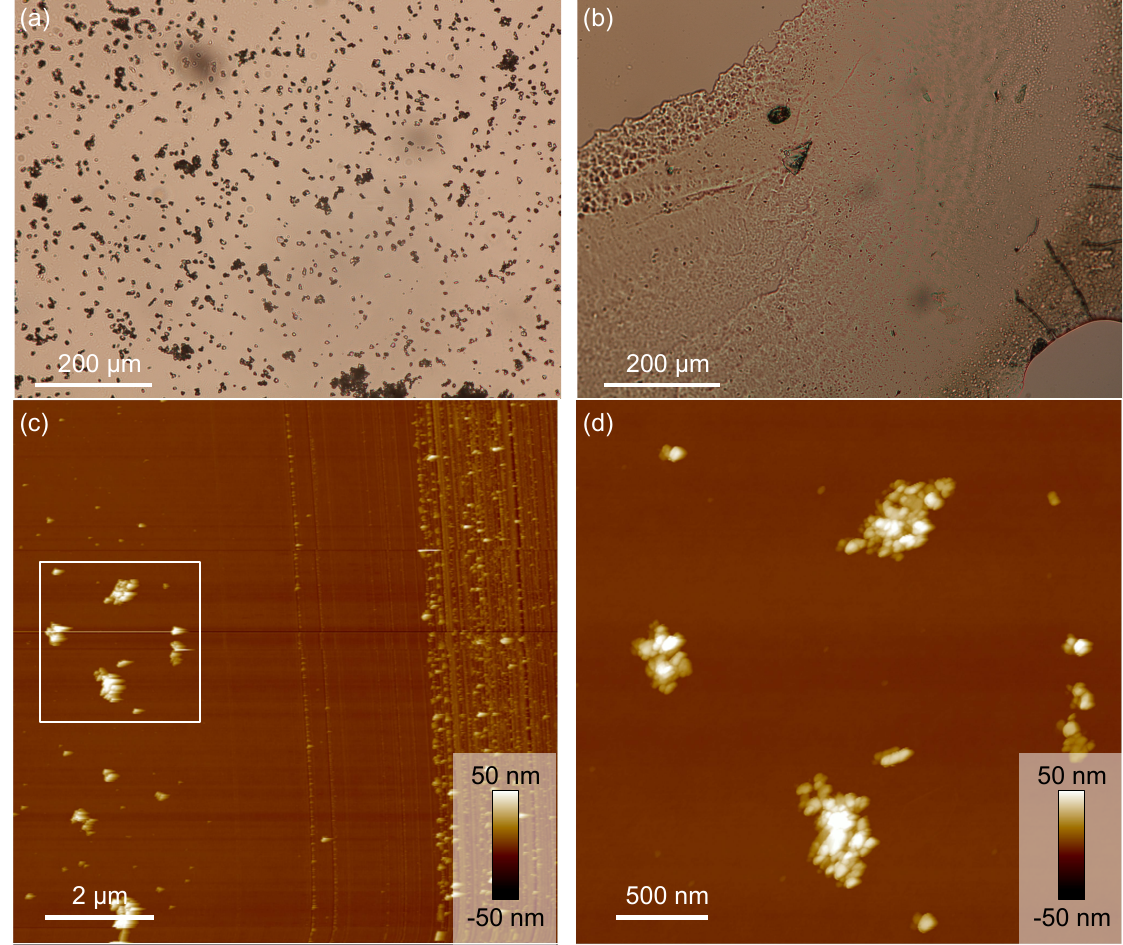}  
	\caption{(a) Optical image of a sample with 3~$\mu$m diamonds deposited on glass at 20x magnification. While some agglomeration is visible, the individual diamond particles can also be resolved.
    (b) Optical image of a sample with 40~nm diamonds deposited on glass at 20x magnification.
    (c) Surface topography of a sample with 40~nm diamonds deposited on glass, imaged with AFM. The area on the left shows well-separated structures, with average distances between them larger than the focused laser beam diameter.
    (d) An AFM image of the white box from panel (c). Larger bright flakes with a diameter of several hundred nm are nanodiamond aggregates. We also find several individual nanodiamonds with $d\sim 40~\mathrm{nm}$.
    }
	\label{fig:sampleImages}
\end{figure*}

Both optical microscopy and atomic force microscopy (AFM) were used to monitor the sample quality. When depositing $3 ~\mathrm{\mu m}$ diamonds, individual diamond particles can be well resolved under the optical microscope using 20x magnification (Fig. \ref{fig:sampleImages} (a)). In contrast, optical microscopy alone is insufficient to assess the quality of deposition when using 40~nm diamonds (Fig. \ref{fig:sampleImages} (b)). For these samples, we used an AFM to image the sample surface wits sub-$\mu$m resolution. While substrate defects, such as scratches on the glass (seen on the right side of Fig. \ref{fig:sampleImages} (c)) lead to significant aggregation of nanodiamonds, we also find large areas where individual features are well-separated. A high-resolution image,  Fig. \ref{fig:sampleImages} (d), reveals that many such features are hundreds of nm in size and are thus aggregates of the 40~nm nanodiamonds. Individual $d\sim 40~\mathrm{nm}$ can still be observed.

\pagebreak
\section{Fitting of relaxation curves}

\begin{figure*}[htbp]
	\centering
	\includegraphics[width=0.8\textwidth]{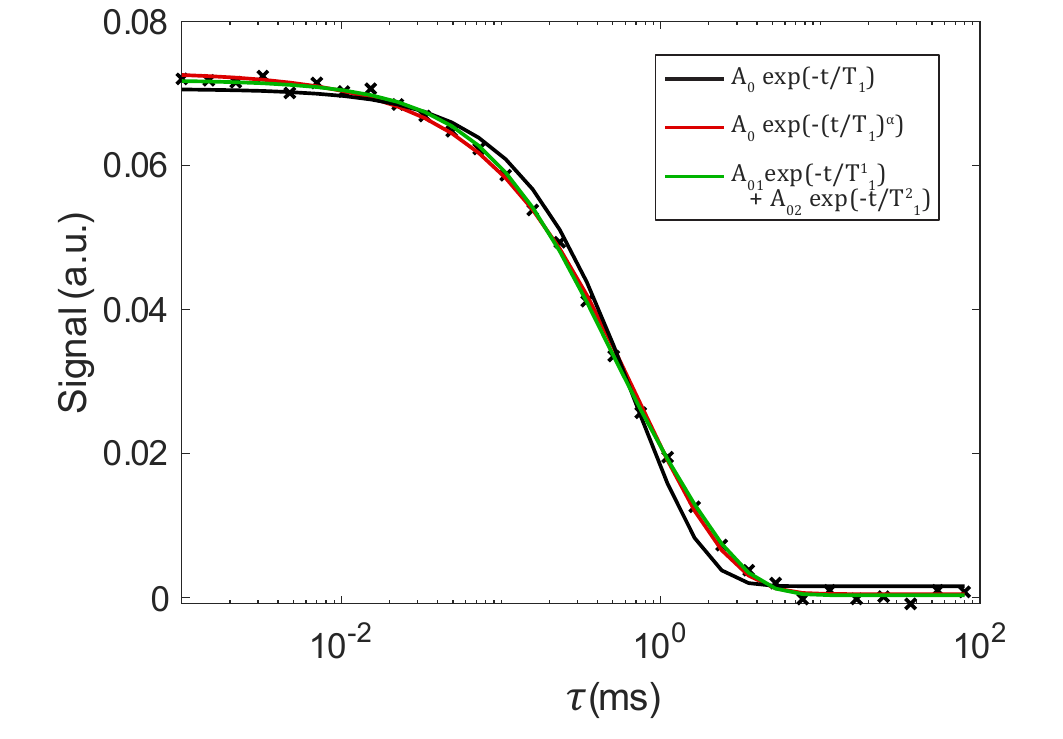}  
	\caption{Comparison of various relaxation functions fitted to a typical relaxation curve. The single-exponential curve (solid black line) does not follow the data well, and more complex fit functions need to be considered. The stretched exponential (red) and the two-component exponential (green) fit functions yield an almost identical result. Due to random distribution of NV center locations within a nanodiamond, the two-component is difficult to justify, and thus stretched exponential was used for all measurements.}
	\label{fig:curveFits}
\end{figure*}

To find the most appropriate fitting curve for the data acquired during the relaxometry measurement, we fitted the three candidate functions: a single-exponential curve of the form $A= A_0e^{-\tau/T_1}$, a two-exponential curve of the form $A= A_{01}e^{-\tau/T^1_1} + A_{02}e^{-\tau/T^2_1}$ and the stretched exponential curve $A=A_0 e^{(-\tau/T_1)^\alpha}$. The data taken on $3 ~\mathrm{\mu m}$ diamonds on glass substrate is presented in Fig. \ref{fig:curveFits} alongside with the three fitting curves. The single-exponential curve (black line) notably deviates from the data point in the region between $\tau = 100 ~\mathrm{\mu s}$ and $\tau = 10 ~\mathrm{ms}$. Both the two-exponential (green) and the stretched exponential (red) curves fit the data well in the whole range of values of $\tau$. To quantify the difference, we calculated the mean squared error (MSE) for each of the three curves. For the single-exponential curve, $\mathrm{MSE}_\mathrm{single} = 1.1 \cdot 10^{-4}$ is an order of magnitude larger than the $\mathrm{MSE}_\mathrm{two} = 1.1 \cdot 10^{-5}$ and $\mathrm{MSE}_\mathrm{stretch} = 1.6 \cdot 10^{-5}$ for two-exponential and stretched curve, respectively.\par
While the two-exponential curve yields the best overall goodness-of-fit parameter, the long relaxation component is consistently very large, $T_1 > 1.2 ~\mathrm{ms}$, with a large fitting uncertainty. Because the data does not clearly show two different timescales, we opted for using the stretched exponential fit, which has the advantage of having fewer free parameters.

